\documentclass[%
 reprint,
 amsmath,amssymb,
 aps,
 prl,
floatfix
]{revtex4-1}

\usepackage{graphicx}
\usepackage{dcolumn}
\usepackage{bm}

\usepackage{color}
\usepackage{textcomp}
\usepackage{mathtools}

\DeclarePairedDelimiterX\braket[2]{\langle}{\rangle}{#1 \delimsize\vert #2}

\begin{document} 
\title{Metallicity in Ultra-Thin Oxygen-Deficient SrTiO$_3$ Thin Films}

\author{P. P. Balakrishnan}
\email{purnimab@stanford.edu}
\affiliation{%
 Department of Physics and Geballe Laboratory for Advanced Materials, Stanford University, Stanford, CA
}%
\author{U. S. Alaan, M. T. Gray}
\affiliation{%
 Department of Materials Science and Engineering and Geballe Laboratory for Advanced Materials, Stanford University, Stanford, CA
}%
\author{Y. Suzuki}
\affiliation{%
 Department of Applied Physics and Geballe Laboratory for Advanced Materials, Stanford University, Stanford, CA
}%


\begin{abstract}
We report on the observation of metallic behavior in thin films of oxygen-deficient SrTiO$_3$ -- down to 9 unit cells -- when coherently strained on (001) SrTiO$_3$ or DyScO$_3$-buffered (001) SrTiO$_3$ substrates. These films have carrier concentrations of up to 2$\times$10$^{22}$ cm$^{-3}$ and mobilities of up to 19,000 cm$^2$/V$\cdot$s at 2 K. There exists a non-conducting layer in our SrTiO$_{3-\delta}$ films that is larger in films with lower carrier concentrations. This non-conducting layer can be attributed to a surface depletion layer due to a Fermi level pinning potential. The depletion width, transport, and structural properties are not greatly affected by the insertion of a DyScO$_3$ buffer between the SrTiO$_3$ film and SrTiO$_3$ substrate.
\end{abstract}

\maketitle

Among complex oxide materials, SrTiO$_3$ (STO) has been extensively studied for many decades, both in bulk and thin film forms. STO is a band insulator with a bandgap of 3.2 eV \cite{Tufte1967, Capizzi1970, Mattheiss1972, VanBenthem2001, Velev2005}. Its nearly cubic perovskite crystal structure, controllable surface termination \cite{Kawasaki1994, Koster1998}, and insulating behavior make it an ideal substrate for complex oxide thin film growth. Due to its widespread use, many advances in crystal growth have been made to improve crystal quality and reduce impurity levels in STO.

STO exhibits electronic properties which can be easily modulated through doping, defects, or structural modification. Metallicity has been induced in the conventionally band-insulating STO in many ways: by doping, for example with Nb \cite{Tufte1967} or La \cite{Son2010a, Suzuki1996}; by exciting photocarriers \cite{Kozuka2008}; and by oxygen depletion \cite{Schooley1964, Schooley1965, Frederikse1967a, Tufte1967, Koonce1967, Lee1971, Henrich1978}. Metallicity has also been observed at the surfaces of vacuum-cleaved \cite{Santander-Syro2011} and Ar$^{+}$-irradiated STO single crystals \cite{Kan2005}, which has been attributed to oxygen depletion at the surface.

Studies on oxygen-deficient STO films have been hampered by an inability to accurately quantify oxygen stoichiometry in conducting films. For example, Ohtomo et al.\ have varied the ambient oxygen pressure and temperature during STO thin film deposition to modulate transport behavior between metallic and insulating states \cite{Ohtomo2007a}. However, in these experiments, the relative oxygen content was deduced from carrier concentration values, but the actual oxygen stoichiometry in these films could not be directly measured.

Other groups have explored the role of coherent strain on oxygen-deficient STO films. A recent study by Huang et al.\ showed that even when STO is doped with carriers, lattice strain can dramatically suppress metallic conduction \cite{Huang2014}. They explain the suppression of metallicity in SrTiO$_{3-\delta}$ films under tensile and compressive strain in terms of electronic structure modification that reduces the density of states at the Fermi level. This limits the choice of substrate for the stabilization of metallic STO films to STO single crystals.

When depositing oxygen-deficient films, an oxygen-poor atmosphere and elevated temperatures are generally required prior to and during film growth. These conditions can also induce oxygen vacancies in the underlying oxide substrate. The presence of a two-dimensional electron gas at the surface of a vacuum-cleaved STO crystal is evidence that oxygen vacancies can be induced at the surface of a crystal under the right conditions \cite{Santander-Syro2011}. Since metallicity in these films requires homoepitaxial growth, care must be taken to distinguish the oxygen-deficient STO layer from the STO substrate.

In addition to oxygen vacancy defects and lattice strain, surface states due to disorder, dislocations, or dangling bonds can give rise to a surface potential that is pinned, resulting in a depletion of carriers from the surface of the STO films \cite{Ohtomo2004}. In order to understand conduction in STO films, the effects of lattice distortions, surface depletion and oxygen diffusion must all be taken into account. 

In this paper, we demonstrate that oxygen-deficient STO films can exhibit metallic behavior down to 9 unit cells thick -- with carrier concentrations as high as 2$\times$10$^{22}$~cm$^{-3}$ and mobilities as high as 19,000 cm$^2$/V$\cdot$s at 2~K -- when grown on single-crystal or buffered (001) STO substrates. The metallicity can be correlated with an expansion in the out-of-plane lattice parameter by 0.6\%. However even small lattice distortions induced by film growth directly on perovskite-structure substrates with different lattice parameters give rise to insulating behavior in oxygen-deficient STO films. When a thin, coherently strained buffer layer of DSO is inserted between the oxygen-deficient STO film and STO substrate, metallicity is observed as in homoepitaxial films. This suggests that strain, not a difference in chemical potential, is in fact a primary driver for control of metallicity in these films. Thickness-dependent transport measurements indicate that there is a non-conducting layer in homoepitaxial STO films and STO films grown on DSO-buffered STO substrates. The dramatic dependence of this non-conducting layer on oxygen growth pressure suggests that it is due to surface depletion caused by a surface pinning potential.

All films were grown by pulsed laser deposition, using a KrF laser ($\lambda=248$ nm) operating at 1 Hz with a laser fluence of 0.7 J/cm$^2$. Oxygen-deficient STO films were grown using a sintered SrTiO$_3$ ceramic target at 750\textdegree C and a background pressure of 5$\times$10$^{-6}$ Torr. Films with a range of thicknesses (2--55 nm) were deposited onto TiO$_2$-terminated (001) STO substrates, both with and without a 1.5 nm DSO buffer layer. We grew the buffer layer on the STO substrate at 750\textdegree C in 5$\times$10$^{-4}$ Torr O$_2$, and then subsequently lowered the growth pressure without exposure to atmosphere before depositing the oxygen-deficient STO film. Thick DSO control films were grown in vacuum and in oxygen atmospheres of up to 5$\times$10$^{-4}$ Torr O$_2$ in order to confirm that the transport behavior of the DSO films was insulating. Oxygen-deficient STO films were also grown on (001) LaAlO$_3$ (LAO), (001) (LaAlO$_3$)$_{0.3}$(Sr$_2$AlTaO$_6$)$_{0.7}$ (LSAT), and (110) DyScO$_3$ (DSO) single crystal substrates. For comparison, oxygen-deficient STO films were also grown in atmospheres of up to 5$\times$10$^{-5}$ Torr O$_2$ on STO substrates. 

Both structure and electrical transport were characterized by a variety of methods. Atomic force microscopy (AFM) was performed using a Veeco Dimension 3100 to measure surface roughness of the films. $\theta$-2$\theta$ and $\omega$ scans were measured by X-ray diffraction (XRD) on a PANalytical X'Pert Materials Research Diffractometer to investigate strain states and crystallinity. Thickness fringes in the XRD scans were used to determine STO film thicknesses for films greater than 30 nm, while X-ray reflectivity was used to measure DSO film thicknesses. For ultra-thin films, thicknesses were estimated by linearly extrapolating growth time from longer depositions. In- and out-of-plane lattice parameters were extracted from reciprocal space maps (RSM) performed on the same instrument.

Transport measurements were performed in magnetic fields of up to 9 T orthogonal to the film plane and at temperatures between 2 and 300 K using a Quantum Design Physical Property Measurement system. Using Al wire, conductive samples were measured in a van der Pauw configuration, while insulating samples were measured in a Hall bar configuration.

\begin{figure}
\centering
\includegraphics[width=.5\textwidth]{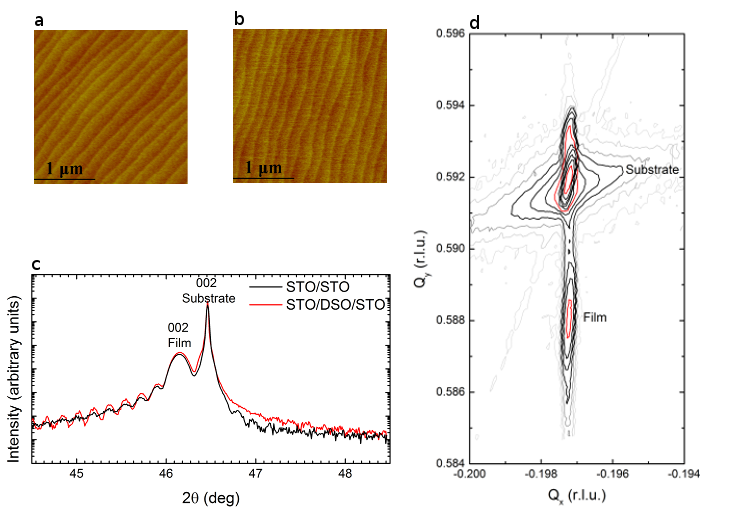}
\caption{\label{fig:XRD}
AFM shows clear atomic step terraces in oxygen-deficient STO films on both (a) bare and (b) DSO-buffered STO substrates, indicating excellent film quality with atomically smooth surfaces (RMS roughness $\sim$0.15 nm) comparable to that of the bare substrate. (c) $\theta$-2$\theta$ scan of the STO 002 peak of $\sim$53 nm homoepitaxial and DSO-buffered films. The film peak position results from an expansion of 0.6\% in the out-of-plane direction, while film thicknesses can be calculated from the periodicity of the reflection fringes, which are clearer in the buffered films. (d) RSM around the STO $\bar{1}03$ peak of the homoepitaxial film in (c), showing that the film is coherently strained to the substrate, and showing the same out-of-plane expansion as in (c).}
\end{figure}

Oxygen-deficient STO films -- grown both directly on (001) STO substrates and on DSO-buffered STO substrates -- exhibit epitaxy and excellent crystalline quality (Figure \ref{fig:XRD}). In $\theta$-2$\theta$ scans of the buffered and unbuffered films, the presence of oxygen-deficient STO films can be distinguished quite readily. Only the \{001\} film peaks are present, and they are separated from the STO substrate peaks, as shown in Figure \ref{fig:XRD}c. The STO film peaks reveal an expanded out-of-plane lattice parameter due to oxygen vacancies of approximately 3.93 \AA\ (compared to 3.905 \AA\ in the bulk) regardless of the presence or absence of the buffer layer. Thicknesses of the oxygen-deficient STO films were deduced from the spacing of the satellite peaks of the 002 film peak. Rocking curve $\omega$ scans of the 002 film peaks indicate excellent crystallinity of the films, with full-width-at-half-maximum (FWHM) in the range of $\Delta$$\omega$ = .04\textdegree--.09\textdegree. Since the DSO buffer layers are only 3--4 unit cells thick, their presence could not be uniquely identified in the XRD data. However, scans of thicker DSO films grown on STO substrates show the orthorhombic \{110\} family of peaks as expected.

Lattice distortions were further studied with reciprocal space maps (RSM) around the STO $\bar{1}03$ peak. RSMs were measured on samples 35--55 nm thick. The films are coherently strained to the substrate and are expanded out-of-plane to 3.93~\AA\ from the bulk value of 3.905~\AA\ (Figure \ref{fig:XRD}d), in agreement with the $\theta$-2$\theta$ scans. These in- and out-of-plane lattice parameters correspond to a volume expansion of around 0.6\% for both homoepitaxial and buffered films. This volume expansion is expected in oxygen-deficient STO and is similar to that seen by Ohtomo et al.\ \cite{Ohtomo2007a}. Regardless of the presence of the DSO buffer layer, the metallic STO films exhibit epitaxy and similarly excellent crystallinity.

For comparison, we studied oxygen-deficient STO films grown epitaxially on tetragonal (001) LSAT, orthorhombic (110) DSO, and trigonal (110) LAO single-crystal substrates; these substrates have a lattice mismatch of $-0.9\%$, $+1.0\%$, and $-2.9\%$ from bulk STO, respectively. STO films are coherently strained on LSAT and DSO, while films on LAO, which has a larger lattice mismatch with STO, are partially relaxed, as measured by reciprocal space mapping. In these heteroepitaxial films, rocking curve $\omega$ scans of the 002 film peaks indicate excellent crystallinity, with FWHM in the range of $\Delta$$\omega$ = .04\textdegree--.06\textdegree. These $\Delta$$\omega$ values are comparable to the STO films grown on STO or DSO-buffered STO substrates, indicating that the homoepitaxial and heteroepitaxial films have similar crystallinities.

\begin{figure}
\centering
\includegraphics[width=0.5\textwidth]{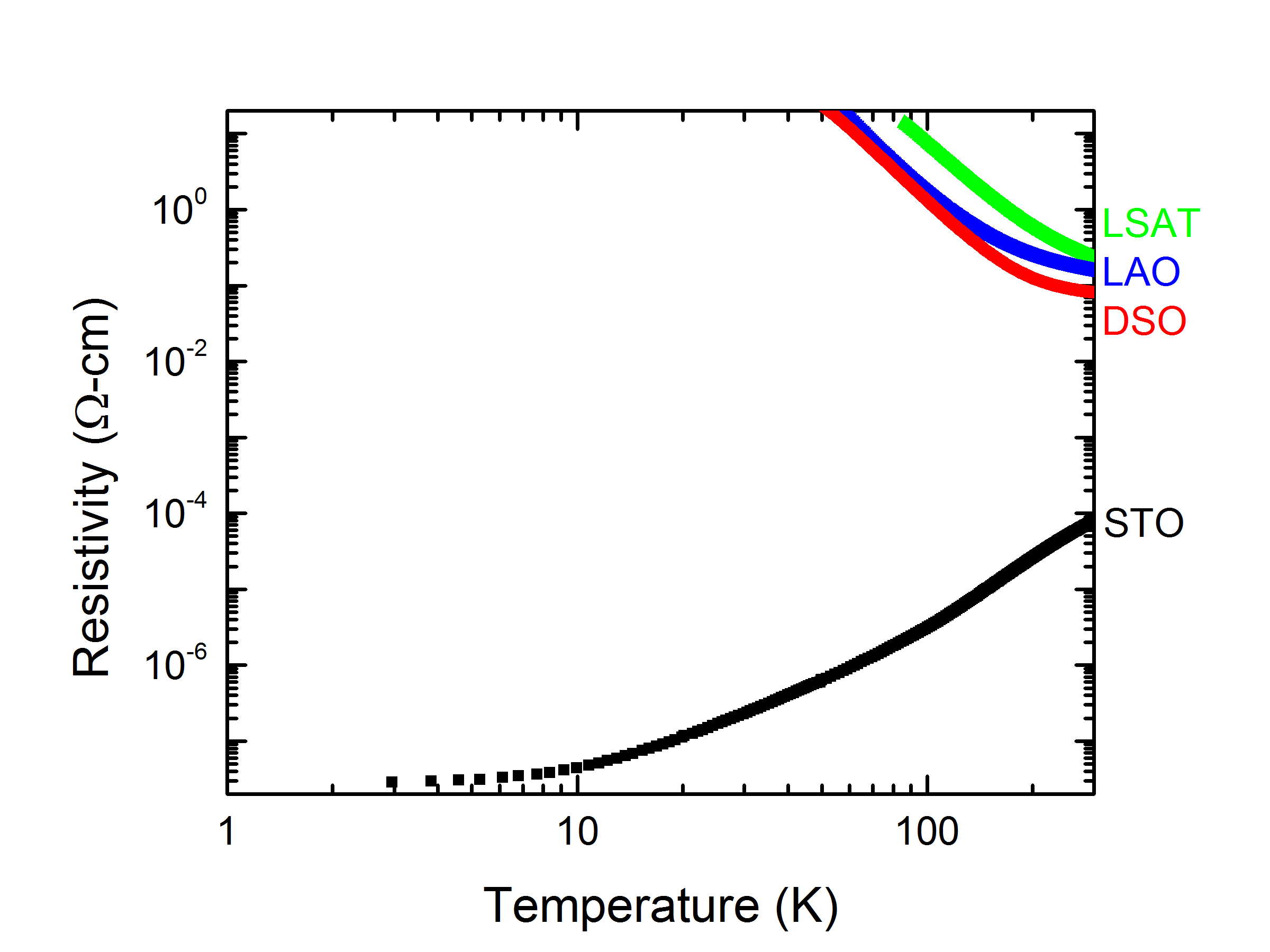}
\caption{\label{fig:resothersub} Temperature dependence of resistivity of oxygen-deficient STO strained on various substrates: (001) LAO, (001) LSAT, (001) STO, and (110) DSO. Films on STO are metallic, while films on DSO, LSAT, and LAO are insulating.}
\end{figure}

Transport measurements indicate that metallic behavior is observed only in homoepitaxial and DSO-buffered STO films, but not in heteroepitaxial STO films grown on LAO, LSAT, or DSO substrates (Figure \ref{fig:resothersub}). Even at room temperature, the resistivities of the heteroepitaxial, insulating STO films are more than three orders of magnitude higher than those of homoepitaxial, metallic STO films. This is consistent with results from Huang et al.\ \cite{Huang2014}, which showed that oxygen-deficient films grown under slightly different growth conditions on DSO, LAO, and LSAT substrates are insulating \cite{Huang2014}. Together, these results indicate the importance of the STO crystal structure in stabilizing the metallic ground state in oxygen-deficient STO films.

\begin{figure}
\centering
\includegraphics[width=0.5\textwidth]{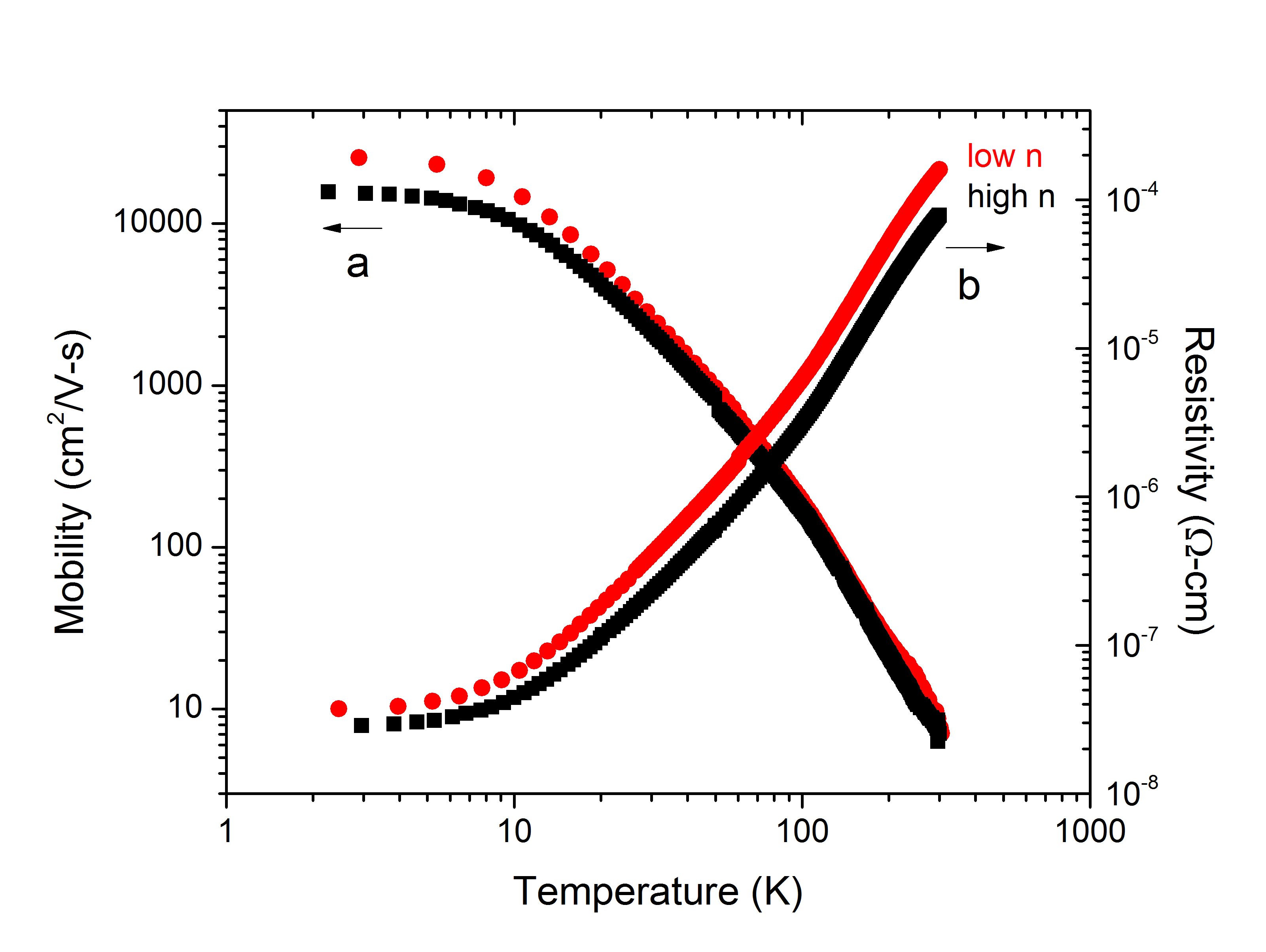}
\caption{Temperature dependence of the mobility (a) and resistivity (b) of oxygen-deficient STO films grown at  different pressures. Films were grown at 5$\times$10$^{-6}$ Torr (base pressure) and 6.5$\times$10$^{-6}$ Torr O$_2$, resulting in carrier concentrations $n$ of 1.1$\times$10$^{22}$~cm$^{-3}$ and 5$\times$10$^{21}$~cm$^{-3}$, respectively. The STO films are $\sim$45 nm thick.}
\label{fig:RvsT}
\end{figure}

The metallic temperature dependence of the sheet resistance is very similar for homoepitaxial and DSO buffered STO samples of comparable thicknesses. The sheet resistance follows a power law dependence of $\sim$T$^3$ at temperatures above the cubic-to-orthorhombic structural transition of bulk STO around 105 K (Figure \ref{fig:RvsT}b). The sheet resistance scales with thickness for both homoepitaxial and DSO buffered samples, indicating that the primary source of metallic conduction is in fact the STO film, not the bulk DSO film or the DSO/STO interface.

Hall effect measurements on these metallic samples show that the carriers are indeed electrons with carrier concentrations that are largely independent of temperature as expected for metals. The carrier concentrations for our films are typically 8$\times$10$^{21}$--1.6$\times$10$^{22}$ cm$^{-3}$ regardless of thickness. These high carrier concentrations indicate highly doped STO. Even with the high carrier concentration values, the mobility values are as high as 19,000 cm$^2$/V$\cdot$s at low temperature and typically 5--7 cm$^2$/V$\cdot$s at room temperature. Similar carrier concentrations and electron mobilities have been observed by Ohtomo et al.\ \cite{Ohtomo2007a}.

The temperature dependence of mobility of oxygen-deficient STO films can be described in terms of two regimes (Figure \ref{fig:RvsT}a). At low temperatures, the mobility flattens out and can be attributed to defect scattering. At higher temperatures, the mobility can be described by a power law, indicating the influence of phonon-scattering.

For comparison, homoepitaxial STO films were also grown in slightly higher ambient oxygen pressures of 6.5$\times$10$^{-6}$ Torr and 5$\times$10$^{-5}$ Torr. Films grown in 6.5$\times$10$^{-6}$ Torr still show metallic behavior (Figure \ref{fig:RvsT}), but have lower carrier concentration values on the order of 5$\times$10$^{21}$ cm$^{-3}$ accompanied by slightly higher mobility values of 25,000 cm$^2$/V$\cdot$s. Films grown in 5$\times$10$^{-5}$ Torr of oxygen are no longer metallic, indicating that the higher oxygen pressures are enough to suppress the formation of oxygen vacancies in the STO films.

\begin{figure}
\centering
\includegraphics[width=0.5\textwidth]{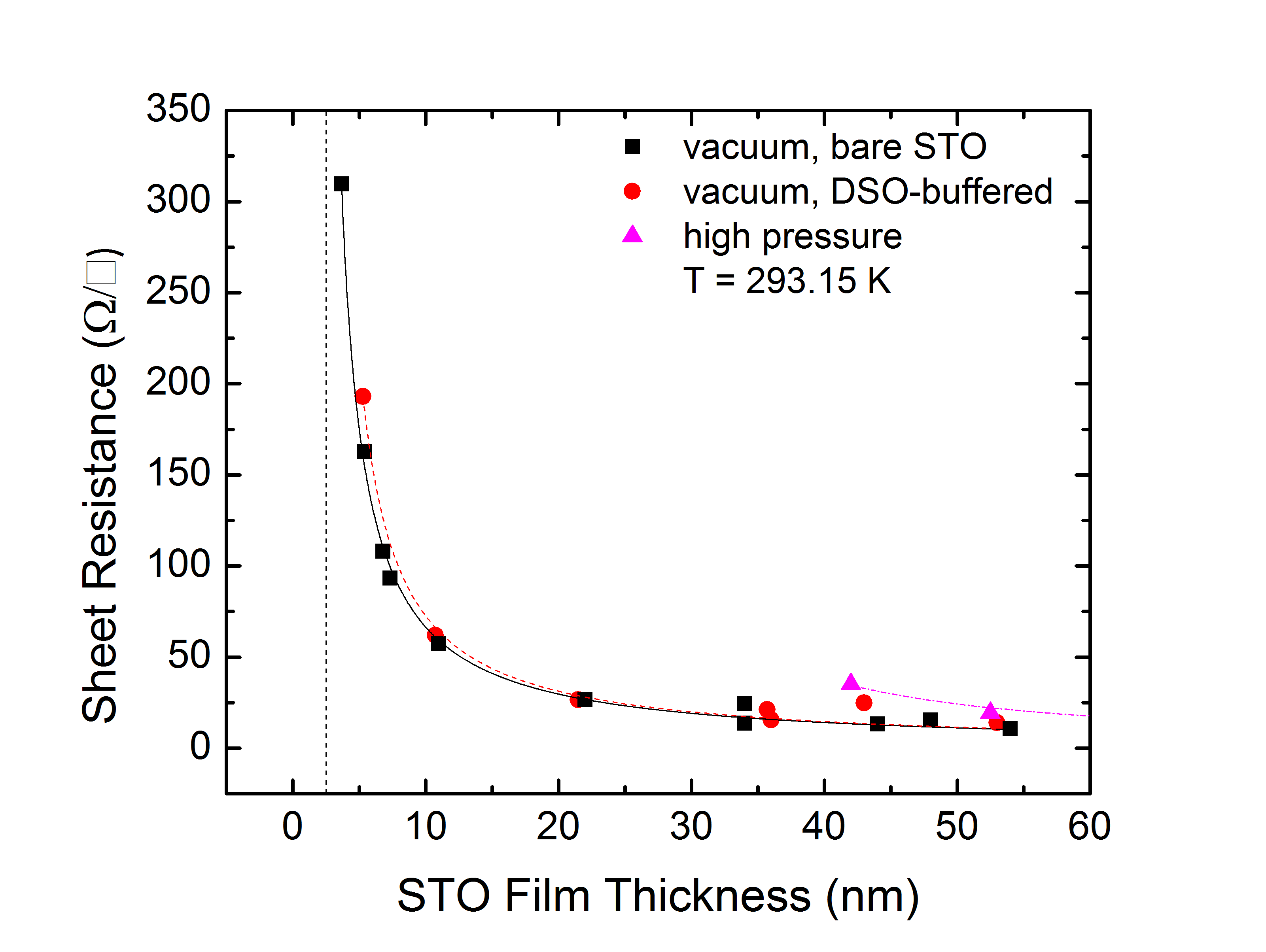}
\caption{\label{fig:Rvst}Thickness dependence of the room-temperature sheet resistance, showing an inverse relationship with a thickness offset, indicating a non-conducting layer of $\sim$2 nm for films grown in vacuum, and $\sim$20 nm for films grown in 6.5$\times$10$^{-6}$ Torr O$_2$.}
\end{figure}

The room-temperature sheet resistance of all homoepitaxial STO films and STO films on DSO-buffered STO substrates is found to scale inversely with thickness, with a thickness offset of approximately 2 nm as shown in Figure \ref{fig:Rvst}. Correspondingly, when the 2D carrier concentration is plotted against the STO film thickness, there is a linear dependence with the intercept on the horizontal axis corresponding to the thickness of a non-conducting layer at either the surface or the interface with the underlying STO or DSO-buffered STO. This layer is approximately 5 unit cells or 2 nm and largely independent of temperature in both the homoepitaxial and DSO-buffered samples. For films grown under higher oxygen pressures with lower carrier concentration values of 5$\times$10$^{21}$ cm$^{-3}$, this non-conducting layer is estimated to be almost an order of magnitude higher at around 20 nm.

This non-conducting layer within the film may occur at either the film surface or its interface with the substrate, or some combination of the two. Some degree of microstructural disorder could occur at the interface between the film and the substrate or buffer layer, thus giving rise to a non-conducting layer. Alternatively, the non-conducting layer could be at the surface of the film due to surface depletion.

Surface states from disorder, defects, or dangling bonds can give rise to a surface potential that is pinned at a specific value above the conduction band, resulting in a depletion of carriers from the surface of STO films \cite{Ohtomo2004}. This surface depletion is thought to be particularly significant for low carrier concentrations which, when combined with the large dielectric constant of STO, can give rise to large depletion widths upwards of hundreds of nanometers \cite{Hyeokunpublished}. 

If we assume a simplistic semiconductor band model, the depletion width can be written in the form $W = \left(2\epsilon V_B/q n\right)^{1/2}$ where $\epsilon$ is the absolute permittivity of STO, V$_B$ is the pinning potential, $q$ is the elementary charge, and $n$ is the carrier concentration \cite{Chandra1979}. For a depletion layer thickness of 2 nm at room temperature and carrier concentration of 8$\times$10$^{21}$ cm$^{-3}$, we estimate the pinning potential to be 0.93 eV, which is similar to that measured in La-doped STO films \cite{Ohtomo2004}. However, for the same pinning potential and a carrier concentration of 5$\times$10$^{21}$ cm$^{-3}$ (corresponding to films grown at the slightly higher ambient pressure), we would expect a depletion width of only 2.5 nm, rather than 20 nm. 

We would not expect such a small increase of 1.5$\times$10$^{-6}$ Torr in oxygen growth pressure to change the disorder at the interface so dramatically so as to increase the non-conducting layer thickness by almost an order of magnitude; AFM and rocking curves of these films show the same surface roughness and crystallinity as for films grown in vacuum. In fact, we would expect an increased oxygen pressure during growth to improve oxygen stoichiometry in film, lowering the number of defects and scattering, and therefore to have the opposite trend.

In addition, this semiconductor model assumes that (i) the depletion width is finite, with no mobile carriers, (ii) the conducting region starts abruptly, with uniform carrier concentration, and (iii) the dielectric constant is linear and therefore does not depend on the field or pinning potential itself. These assumptions do not hold for oxygen-deficient STO \cite{Neville1972}, suggesting that the true behavior is more complicated. However because lower carrier concentrations increase the thickness of the non-conducting layer in our samples, it may be at least partly attributed to a depletion layer due to Fermi level pinning at the surface instead of a disordered layer at the interface.

It is interesting to note that there have been reports of a significant depletion width in doped STO that may prevent it from being exploited in some oxide electronics devices \cite{Ohtomo2004, Hyeokunpublished}. Our results on oxygen-deficient STO films represent a higher carrier concentration range than previous reports of STO films, thereby minimizing (although not eliminating) the surface depletion layer and possibly improving the chances for incorporation of oxygen-vacancy-doped STO into oxide electronic devices. 

In summary, we demonstrate the growth of oxygen-deficient STO films on STO substrates and DSO-buffered STO substrates with carrier concentration and mobility values as high as 2$\times$10$^{22}$ cm$^{-3}$ and 19,000 cm$^2$/V$\cdot$s respectively. Metallic behavior is observed down to 9 unit cells thick below which it is insulating. The non-conducting layer can be attributed to a surface depletion layer that is inversely correlated with carrier concentration. By increasing carrier concentrations and thereby decreasing surface depletion widths, metallic STO layers may very well be exploited in oxide electronic devices.

We would like to thank Hyeok Yoon, Ted Sanders, Michael Veit, and Harold Hwang for insightful discussions, and Arturas Vailionis for assistance with X-ray diffraction. This work was funded by the the National Science Foundation under DMR-1402685. X-ray diffraction measurements were performed at the Stanford Nano Shared Facilities at Stanford University.

\bibliography{main.bib}

\begin{thebibliography}{24}%
\makeatletter
\providecommand \@ifxundefined [1]{%
 \@ifx{#1\undefined}
}%
\providecommand \@ifnum [1]{%
 \ifnum #1\expandafter \@firstoftwo
 \else \expandafter \@secondoftwo
 \fi
}%
\providecommand \@ifx [1]{%
 \ifx #1\expandafter \@firstoftwo
 \else \expandafter \@secondoftwo
 \fi
}%
\providecommand \natexlab [1]{#1}%
\providecommand \enquote  [1]{``#1''}%
\providecommand \bibnamefont  [1]{#1}%
\providecommand \bibfnamefont [1]{#1}%
\providecommand \citenamefont [1]{#1}%
\providecommand \href@noop [0]{\@secondoftwo}%
\providecommand \href [0]{\begingroup \@sanitize@url \@href}%
\providecommand \@href[1]{\@@startlink{#1}\@@href}%
\providecommand \@@href[1]{\endgroup#1\@@endlink}%
\providecommand \@sanitize@url [0]{\catcode `\\12\catcode `\$12\catcode
  `\&12\catcode `\#12\catcode `\^12\catcode `\_12\catcode `\%12\relax}%
\providecommand \@@startlink[1]{}%
\providecommand \@@endlink[0]{}%
\providecommand \url  [0]{\begingroup\@sanitize@url \@url }%
\providecommand \@url [1]{\endgroup\@href {#1}{\urlprefix }}%
\providecommand \urlprefix  [0]{URL }%
\providecommand \Eprint [0]{\href }%
\providecommand \doibase [0]{http://dx.doi.org/}%
\providecommand \selectlanguage [0]{\@gobble}%
\providecommand \bibinfo  [0]{\@secondoftwo}%
\providecommand \bibfield  [0]{\@secondoftwo}%
\providecommand \translation [1]{[#1]}%
\providecommand \BibitemOpen [0]{}%
\providecommand \bibitemStop [0]{}%
\providecommand \bibitemNoStop [0]{.\EOS\space}%
\providecommand \EOS [0]{\spacefactor3000\relax}%
\providecommand \BibitemShut  [1]{\csname bibitem#1\endcsname}%
\let\auto@bib@innerbib\@empty
\bibitem [{\citenamefont {Tufte}\ and\ \citenamefont
  {Chapman}(1967)}]{Tufte1967}%
  \BibitemOpen
  \bibfield  {author} {\bibinfo {author} {\bibfnamefont {O.~N.}\ \bibnamefont
  {Tufte}}\ and\ \bibinfo {author} {\bibfnamefont {P.~W.}\ \bibnamefont
  {Chapman}},\ }\href {\doibase 10.1103/PhysRev.155.796} {\bibfield  {journal}
  {\bibinfo  {journal} {Physical Review}\ }\textbf {\bibinfo {volume} {155}},\
  \bibinfo {pages} {796} (\bibinfo {year} {1967})}\BibitemShut {NoStop}%
\bibitem [{\citenamefont {Capizzi}\ and\ \citenamefont
  {Frova}(1970)}]{Capizzi1970}%
  \BibitemOpen
  \bibfield  {author} {\bibinfo {author} {\bibfnamefont {M.}~\bibnamefont
  {Capizzi}}\ and\ \bibinfo {author} {\bibfnamefont {A.}~\bibnamefont
  {Frova}},\ }\href {\doibase 10.1103/PhysRevLett.25.1298} {\bibfield
  {journal} {\bibinfo  {journal} {Physical Review Letters}\ }\textbf {\bibinfo
  {volume} {25}},\ \bibinfo {pages} {1298} (\bibinfo {year}
  {1970})}\BibitemShut {NoStop}%
\bibitem [{\citenamefont {Mattheiss}(1972)}]{Mattheiss1972}%
  \BibitemOpen
  \bibfield  {author} {\bibinfo {author} {\bibfnamefont {L.~F.}\ \bibnamefont
  {Mattheiss}},\ }\href {\doibase 10.1103/PhysRevB.6.4718} {\bibfield
  {journal} {\bibinfo  {journal} {Physical Review B}\ }\textbf {\bibinfo
  {volume} {6}},\ \bibinfo {pages} {4718} (\bibinfo {year} {1972})}\BibitemShut
  {NoStop}%
\bibitem [{\citenamefont {van Benthem}\ \emph {et~al.}(2001)\citenamefont {van
  Benthem}, \citenamefont {Els{\"{a}}sser},\ and\ \citenamefont
  {French}}]{VanBenthem2001}%
  \BibitemOpen
  \bibfield  {author} {\bibinfo {author} {\bibfnamefont {K.}~\bibnamefont {van
  Benthem}}, \bibinfo {author} {\bibfnamefont {C.}~\bibnamefont
  {Els{\"{a}}sser}}, \ and\ \bibinfo {author} {\bibfnamefont {R.~H.}\
  \bibnamefont {French}},\ }\href {\doibase 10.1063/1.1415766} {\bibfield
  {journal} {\bibinfo  {journal} {Journal of Applied Physics}\ }\textbf
  {\bibinfo {volume} {90}},\ \bibinfo {pages} {6156} (\bibinfo {year}
  {2001})}\BibitemShut {NoStop}%
\bibitem [{\citenamefont {Velev}\ \emph {et~al.}(2005)\citenamefont {Velev},
  \citenamefont {Belashchenko}, \citenamefont {Stewart}, \citenamefont
  {Van~Schilfgaarde}, \citenamefont {Jaswal},\ and\ \citenamefont
  {Tsymbal}}]{Velev2005}%
  \BibitemOpen
  \bibfield  {author} {\bibinfo {author} {\bibfnamefont {J.~P.}\ \bibnamefont
  {Velev}}, \bibinfo {author} {\bibfnamefont {K.~D.}\ \bibnamefont
  {Belashchenko}}, \bibinfo {author} {\bibfnamefont {D.~A.}\ \bibnamefont
  {Stewart}}, \bibinfo {author} {\bibfnamefont {M.}~\bibnamefont
  {Van~Schilfgaarde}}, \bibinfo {author} {\bibfnamefont {S.~S.}\ \bibnamefont
  {Jaswal}}, \ and\ \bibinfo {author} {\bibfnamefont {E.~Y.}\ \bibnamefont
  {Tsymbal}},\ }\href {\doibase 10.1103/PhysRevLett.95.216601} {\bibfield
  {journal} {\bibinfo  {journal} {Physical Review Letters}\ }\textbf {\bibinfo
  {volume} {95}},\ \bibinfo {pages} {2} (\bibinfo {year} {2005})}\BibitemShut
  {NoStop}%
\bibitem [{\citenamefont {Kawasaki}\ \emph {et~al.}(1994)\citenamefont
  {Kawasaki}, \citenamefont {Takahashi}, \citenamefont {Maeda}, \citenamefont
  {Tsuchiya}, \citenamefont {Shinohara}, \citenamefont {Ishiyama},
  \citenamefont {Yonezawa}, \citenamefont {Yoshimoto},\ and\ \citenamefont
  {Koinuma}}]{Kawasaki1994}%
  \BibitemOpen
  \bibfield  {author} {\bibinfo {author} {\bibfnamefont {M.}~\bibnamefont
  {Kawasaki}}, \bibinfo {author} {\bibfnamefont {K.}~\bibnamefont {Takahashi}},
  \bibinfo {author} {\bibfnamefont {T.}~\bibnamefont {Maeda}}, \bibinfo
  {author} {\bibfnamefont {R.}~\bibnamefont {Tsuchiya}}, \bibinfo {author}
  {\bibfnamefont {M.}~\bibnamefont {Shinohara}}, \bibinfo {author}
  {\bibfnamefont {O.}~\bibnamefont {Ishiyama}}, \bibinfo {author}
  {\bibfnamefont {T.}~\bibnamefont {Yonezawa}}, \bibinfo {author}
  {\bibfnamefont {M.}~\bibnamefont {Yoshimoto}}, \ and\ \bibinfo {author}
  {\bibfnamefont {H.}~\bibnamefont {Koinuma}},\ }\href {\doibase
  10.1126/science.266.5190.1540} {\bibfield  {journal} {\bibinfo  {journal}
  {Science}\ }\textbf {\bibinfo {volume} {266}},\ \bibinfo {pages} {1540}
  (\bibinfo {year} {1994})}\BibitemShut {NoStop}%
\bibitem [{\citenamefont {Koster}\ \emph {et~al.}(1998)\citenamefont {Koster},
  \citenamefont {Kropman}, \citenamefont {Rijnders}, \citenamefont {Blank},\
  and\ \citenamefont {Rogalla}}]{Koster1998}%
  \BibitemOpen
  \bibfield  {author} {\bibinfo {author} {\bibfnamefont {G.}~\bibnamefont
  {Koster}}, \bibinfo {author} {\bibfnamefont {B.~L.}\ \bibnamefont {Kropman}},
  \bibinfo {author} {\bibfnamefont {G.~J. H.~M.}\ \bibnamefont {Rijnders}},
  \bibinfo {author} {\bibfnamefont {D.~H.~A.}\ \bibnamefont {Blank}}, \ and\
  \bibinfo {author} {\bibfnamefont {H.}~\bibnamefont {Rogalla}},\ }\href
  {\doibase 10.1063/1.122630} {\bibfield  {journal} {\bibinfo  {journal}
  {Applied Physics Letters}\ }\textbf {\bibinfo {volume} {73}},\ \bibinfo
  {pages} {2920} (\bibinfo {year} {1998})}\BibitemShut {NoStop}%
\bibitem [{\citenamefont {Son}\ \emph {et~al.}(2010)\citenamefont {Son},
  \citenamefont {Moetakef}, \citenamefont {Jalan}, \citenamefont {Bierwagen},
  \citenamefont {Wright}, \citenamefont {Engel-Herbert},\ and\ \citenamefont
  {Stemmer}}]{Son2010a}%
  \BibitemOpen
  \bibfield  {author} {\bibinfo {author} {\bibfnamefont {J.}~\bibnamefont
  {Son}}, \bibinfo {author} {\bibfnamefont {P.}~\bibnamefont {Moetakef}},
  \bibinfo {author} {\bibfnamefont {B.}~\bibnamefont {Jalan}}, \bibinfo
  {author} {\bibfnamefont {O.}~\bibnamefont {Bierwagen}}, \bibinfo {author}
  {\bibfnamefont {N.~J.}\ \bibnamefont {Wright}}, \bibinfo {author}
  {\bibfnamefont {R.}~\bibnamefont {Engel-Herbert}}, \ and\ \bibinfo {author}
  {\bibfnamefont {S.}~\bibnamefont {Stemmer}},\ }\href {\doibase
  10.1038/nmat2750} {\bibfield  {journal} {\bibinfo  {journal} {Nature
  Materials}\ }\textbf {\bibinfo {volume} {9}},\ \bibinfo {pages} {482}
  (\bibinfo {year} {2010})}\BibitemShut {NoStop}%
\bibitem [{\citenamefont {Suzuki}\ \emph {et~al.}(1996)\citenamefont {Suzuki},
  \citenamefont {Bando}, \citenamefont {Ootuka}, \citenamefont {Inoue},
  \citenamefont {Yamamoto}, \citenamefont {Takahashi},\ and\ \citenamefont
  {Nishihara}}]{Suzuki1996}%
  \BibitemOpen
  \bibfield  {author} {\bibinfo {author} {\bibfnamefont {H.}~\bibnamefont
  {Suzuki}}, \bibinfo {author} {\bibfnamefont {H.}~\bibnamefont {Bando}},
  \bibinfo {author} {\bibfnamefont {Y.}~\bibnamefont {Ootuka}}, \bibinfo
  {author} {\bibfnamefont {I.~H.}\ \bibnamefont {Inoue}}, \bibinfo {author}
  {\bibfnamefont {T.}~\bibnamefont {Yamamoto}}, \bibinfo {author}
  {\bibfnamefont {K.}~\bibnamefont {Takahashi}}, \ and\ \bibinfo {author}
  {\bibfnamefont {Y.}~\bibnamefont {Nishihara}},\ }\href {\doibase
  10.1143/JPSJ.65.1529} {\bibfield  {journal} {\bibinfo  {journal} {Journal of
  the Physical Society of Japan}\ }\textbf {\bibinfo {volume} {65}},\ \bibinfo
  {pages} {1529} (\bibinfo {year} {1996})}\BibitemShut {NoStop}%
\bibitem [{\citenamefont {Kozuka}\ \emph {et~al.}(2008)\citenamefont {Kozuka},
  \citenamefont {Susaki},\ and\ \citenamefont {Hwang}}]{Kozuka2008}%
  \BibitemOpen
  \bibfield  {author} {\bibinfo {author} {\bibfnamefont {Y.}~\bibnamefont
  {Kozuka}}, \bibinfo {author} {\bibfnamefont {T.}~\bibnamefont {Susaki}}, \
  and\ \bibinfo {author} {\bibfnamefont {H.~Y.}\ \bibnamefont {Hwang}},\ }\href
  {\doibase 10.1103/PhysRevLett.101.096601} {\bibfield  {journal} {\bibinfo
  {journal} {Physical Review Letters}\ }\textbf {\bibinfo {volume} {101}},\
  \bibinfo {pages} {3} (\bibinfo {year} {2008})}\BibitemShut {NoStop}%
\bibitem [{\citenamefont {Schooley}\ \emph {et~al.}(1964)\citenamefont
  {Schooley}, \citenamefont {Hosler},\ and\ \citenamefont
  {Cohen}}]{Schooley1964}%
  \BibitemOpen
  \bibfield  {author} {\bibinfo {author} {\bibfnamefont {J.~F.}\ \bibnamefont
  {Schooley}}, \bibinfo {author} {\bibfnamefont {W.~R.}\ \bibnamefont
  {Hosler}}, \ and\ \bibinfo {author} {\bibfnamefont {M.~L.}\ \bibnamefont
  {Cohen}},\ }\href@noop {} {\bibfield  {journal} {\bibinfo  {journal} {Phys.
  Rev. Lett.}\ }\textbf {\bibinfo {volume} {12}},\ \bibinfo {pages} {474}
  (\bibinfo {year} {1964})}\BibitemShut {NoStop}%
\bibitem [{\citenamefont {Schooley}\ \emph {et~al.}(1965)\citenamefont
  {Schooley}, \citenamefont {Hosler}, \citenamefont {Ambler}, \citenamefont
  {Becker}, \citenamefont {Cohen},\ and\ \citenamefont
  {Koonce}}]{Schooley1965}%
  \BibitemOpen
  \bibfield  {author} {\bibinfo {author} {\bibfnamefont {J.~F.}\ \bibnamefont
  {Schooley}}, \bibinfo {author} {\bibfnamefont {W.~R.}\ \bibnamefont
  {Hosler}}, \bibinfo {author} {\bibfnamefont {E.}~\bibnamefont {Ambler}},
  \bibinfo {author} {\bibfnamefont {J.~H.}\ \bibnamefont {Becker}}, \bibinfo
  {author} {\bibfnamefont {M.~L.}\ \bibnamefont {Cohen}}, \ and\ \bibinfo
  {author} {\bibfnamefont {C.~S.}\ \bibnamefont {Koonce}},\ }\href {\doibase
  10.1103/PhysRevLett.14.305} {\bibfield  {journal} {\bibinfo  {journal}
  {Physical Review Letters}\ }\textbf {\bibinfo {volume} {14}},\ \bibinfo
  {pages} {305} (\bibinfo {year} {1965})}\BibitemShut {NoStop}%
\bibitem [{\citenamefont {Frederikse}\ and\ \citenamefont
  {Hosler}(1967)}]{Frederikse1967a}%
  \BibitemOpen
  \bibfield  {author} {\bibinfo {author} {\bibfnamefont {H.}~\bibnamefont
  {Frederikse}}\ and\ \bibinfo {author} {\bibfnamefont {W.}~\bibnamefont
  {Hosler}},\ }\href {\doibase 10.1103/PhysRev.161.822} {\bibfield  {journal}
  {\bibinfo  {journal} {Physical Review}\ }\textbf {\bibinfo {volume} {161}},\
  \bibinfo {pages} {822} (\bibinfo {year} {1967})}\BibitemShut {NoStop}%
\bibitem [{\citenamefont {Koonce}\ \emph {et~al.}(1967)\citenamefont {Koonce},
  \citenamefont {Cohen}, \citenamefont {Schooley}, \citenamefont {Hosler},\
  and\ \citenamefont {Pfeiffer}}]{Koonce1967}%
  \BibitemOpen
  \bibfield  {author} {\bibinfo {author} {\bibfnamefont {C.~S.}\ \bibnamefont
  {Koonce}}, \bibinfo {author} {\bibfnamefont {M.~L.}\ \bibnamefont {Cohen}},
  \bibinfo {author} {\bibfnamefont {J.~F.}\ \bibnamefont {Schooley}}, \bibinfo
  {author} {\bibfnamefont {W.~R.}\ \bibnamefont {Hosler}}, \ and\ \bibinfo
  {author} {\bibfnamefont {E.~R.}\ \bibnamefont {Pfeiffer}},\ }\href {\doibase
  10.1103/PhysRev.163.380} {\bibfield  {journal} {\bibinfo  {journal} {Physical
  Review}\ }\textbf {\bibinfo {volume} {163}},\ \bibinfo {pages} {380}
  (\bibinfo {year} {1967})}\BibitemShut {NoStop}%
\bibitem [{\citenamefont {Lee}\ \emph {et~al.}(1971)\citenamefont {Lee},
  \citenamefont {Yahia},\ and\ \citenamefont {Brebner}}]{Lee1971}%
  \BibitemOpen
  \bibfield  {author} {\bibinfo {author} {\bibfnamefont {C.}~\bibnamefont
  {Lee}}, \bibinfo {author} {\bibfnamefont {J.}~\bibnamefont {Yahia}}, \ and\
  \bibinfo {author} {\bibfnamefont {J.~L.}\ \bibnamefont {Brebner}},\ }\href
  {\doibase 10.1103/PhysRevB.3.2525} {\bibfield  {journal} {\bibinfo  {journal}
  {Physical Review B}\ }\textbf {\bibinfo {volume} {3}},\ \bibinfo {pages}
  {2525} (\bibinfo {year} {1971})}\BibitemShut {NoStop}%
\bibitem [{\citenamefont {Henrich}\ \emph {et~al.}(1978)\citenamefont
  {Henrich}, \citenamefont {Dresselhaus},\ and\ \citenamefont
  {Zeiger}}]{Henrich1978}%
  \BibitemOpen
  \bibfield  {author} {\bibinfo {author} {\bibfnamefont {V.~E.}\ \bibnamefont
  {Henrich}}, \bibinfo {author} {\bibfnamefont {G.}~\bibnamefont
  {Dresselhaus}}, \ and\ \bibinfo {author} {\bibfnamefont {H.~J.}\ \bibnamefont
  {Zeiger}},\ }\href {\doibase 10.1103/PhysRevB.17.4908} {\bibfield  {journal}
  {\bibinfo  {journal} {Physical Review B}\ }\textbf {\bibinfo {volume} {17}},\
  \bibinfo {pages} {4908} (\bibinfo {year} {1978})}\BibitemShut {NoStop}%
\bibitem [{\citenamefont {Santander-Syro}\ \emph {et~al.}(2011)\citenamefont
  {Santander-Syro}, \citenamefont {Copie}, \citenamefont {Kondo}, \citenamefont
  {Fortuna}, \citenamefont {Pailh{\`{e}}s}, \citenamefont {Weht}, \citenamefont
  {Qiu}, \citenamefont {Bertran}, \citenamefont {Nicolaou}, \citenamefont
  {Taleb-Ibrahimi}, \citenamefont {Le~F{\`{e}}vre}, \citenamefont {Herranz},
  \citenamefont {Bibes}, \citenamefont {Reyren}, \citenamefont {Apertet},
  \citenamefont {Lecoeur}, \citenamefont {Barth{\'{e}}l{\'{e}}my},\ and\
  \citenamefont {Rozenberg}}]{Santander-Syro2011}%
  \BibitemOpen
  \bibfield  {author} {\bibinfo {author} {\bibfnamefont {A.~F.}\ \bibnamefont
  {Santander-Syro}}, \bibinfo {author} {\bibfnamefont {O.}~\bibnamefont
  {Copie}}, \bibinfo {author} {\bibfnamefont {T.}~\bibnamefont {Kondo}},
  \bibinfo {author} {\bibfnamefont {F.}~\bibnamefont {Fortuna}}, \bibinfo
  {author} {\bibfnamefont {S.}~\bibnamefont {Pailh{\`{e}}s}}, \bibinfo {author}
  {\bibfnamefont {R.}~\bibnamefont {Weht}}, \bibinfo {author} {\bibfnamefont
  {X.~G.}\ \bibnamefont {Qiu}}, \bibinfo {author} {\bibfnamefont
  {F.}~\bibnamefont {Bertran}}, \bibinfo {author} {\bibfnamefont
  {A.}~\bibnamefont {Nicolaou}}, \bibinfo {author} {\bibfnamefont
  {A.}~\bibnamefont {Taleb-Ibrahimi}}, \bibinfo {author} {\bibfnamefont
  {P.}~\bibnamefont {Le~F{\`{e}}vre}}, \bibinfo {author} {\bibfnamefont
  {G.}~\bibnamefont {Herranz}}, \bibinfo {author} {\bibfnamefont
  {M.}~\bibnamefont {Bibes}}, \bibinfo {author} {\bibfnamefont
  {N.}~\bibnamefont {Reyren}}, \bibinfo {author} {\bibfnamefont
  {Y.}~\bibnamefont {Apertet}}, \bibinfo {author} {\bibfnamefont
  {P.}~\bibnamefont {Lecoeur}}, \bibinfo {author} {\bibfnamefont
  {A.}~\bibnamefont {Barth{\'{e}}l{\'{e}}my}}, \ and\ \bibinfo {author}
  {\bibfnamefont {M.~J.}\ \bibnamefont {Rozenberg}},\ }\href {\doibase
  10.1038/nature09720} {\bibfield  {journal} {\bibinfo  {journal} {Nature}\
  }\textbf {\bibinfo {volume} {469}},\ \bibinfo {pages} {189} (\bibinfo {year}
  {2011})}\BibitemShut {NoStop}%
\bibitem [{\citenamefont {Kan}\ \emph {et~al.}(2005)\citenamefont {Kan},
  \citenamefont {Terashima}, \citenamefont {Kanda}, \citenamefont {Masuno},
  \citenamefont {Tanaka}, \citenamefont {Chu}, \citenamefont {Kan},
  \citenamefont {Ishizumi}, \citenamefont {Kanemitsu}, \citenamefont
  {Shimakawa},\ and\ \citenamefont {Takano}}]{Kan2005}%
  \BibitemOpen
  \bibfield  {author} {\bibinfo {author} {\bibfnamefont {D.}~\bibnamefont
  {Kan}}, \bibinfo {author} {\bibfnamefont {T.}~\bibnamefont {Terashima}},
  \bibinfo {author} {\bibfnamefont {R.}~\bibnamefont {Kanda}}, \bibinfo
  {author} {\bibfnamefont {A.}~\bibnamefont {Masuno}}, \bibinfo {author}
  {\bibfnamefont {K.}~\bibnamefont {Tanaka}}, \bibinfo {author} {\bibfnamefont
  {S.}~\bibnamefont {Chu}}, \bibinfo {author} {\bibfnamefont {H.}~\bibnamefont
  {Kan}}, \bibinfo {author} {\bibfnamefont {A.}~\bibnamefont {Ishizumi}},
  \bibinfo {author} {\bibfnamefont {Y.}~\bibnamefont {Kanemitsu}}, \bibinfo
  {author} {\bibfnamefont {Y.}~\bibnamefont {Shimakawa}}, \ and\ \bibinfo
  {author} {\bibfnamefont {M.}~\bibnamefont {Takano}},\ }\href {\doibase
  10.1038/nmat1498} {\bibfield  {journal} {\bibinfo  {journal} {Nature
  Materials}\ }\textbf {\bibinfo {volume} {4}},\ \bibinfo {pages} {816}
  (\bibinfo {year} {2005})}\BibitemShut {NoStop}%
\bibitem [{\citenamefont {Ohtomo}\ and\ \citenamefont
  {Hwang}(2007)}]{Ohtomo2007a}%
  \BibitemOpen
  \bibfield  {author} {\bibinfo {author} {\bibfnamefont {A.}~\bibnamefont
  {Ohtomo}}\ and\ \bibinfo {author} {\bibfnamefont {H.~Y.}\ \bibnamefont
  {Hwang}},\ }\href {\doibase 10.1063/1.2798385} {\bibfield  {journal}
  {\bibinfo  {journal} {Journal of Applied Physics}\ }\textbf {\bibinfo
  {volume} {102}},\ \bibinfo {pages} {083704} (\bibinfo {year}
  {2007})}\BibitemShut {NoStop}%
\bibitem [{\citenamefont {Huang}\ \emph {et~al.}(2014)\citenamefont {Huang},
  \citenamefont {Liu}, \citenamefont {Yang}, \citenamefont {Zeng},
  \citenamefont {Annadi}, \citenamefont {L{\"{u}}}, \citenamefont {Tan},
  \citenamefont {Chen}, \citenamefont {Sun}, \citenamefont {Renshaw~Wang},
  \citenamefont {Zhao}, \citenamefont {Li}, \citenamefont {Zhou}, \citenamefont
  {Han}, \citenamefont {Wu}, \citenamefont {Feng}, \citenamefont {Coey},\ and\
  \citenamefont {Venkatesan}}]{Huang2014}%
  \BibitemOpen
  \bibfield  {author} {\bibinfo {author} {\bibfnamefont {Z.}~\bibnamefont
  {Huang}}, \bibinfo {author} {\bibfnamefont {Z.~Q.}\ \bibnamefont {Liu}},
  \bibinfo {author} {\bibfnamefont {M.}~\bibnamefont {Yang}}, \bibinfo {author}
  {\bibfnamefont {S.~W.}\ \bibnamefont {Zeng}}, \bibinfo {author}
  {\bibfnamefont {A.}~\bibnamefont {Annadi}}, \bibinfo {author} {\bibfnamefont
  {W.~M.}\ \bibnamefont {L{\"{u}}}}, \bibinfo {author} {\bibfnamefont {X.~L.}\
  \bibnamefont {Tan}}, \bibinfo {author} {\bibfnamefont {P.~F.}\ \bibnamefont
  {Chen}}, \bibinfo {author} {\bibfnamefont {L.}~\bibnamefont {Sun}}, \bibinfo
  {author} {\bibfnamefont {X.}~\bibnamefont {Renshaw~Wang}}, \bibinfo {author}
  {\bibfnamefont {Y.~L.}\ \bibnamefont {Zhao}}, \bibinfo {author}
  {\bibfnamefont {C.~J.}\ \bibnamefont {Li}}, \bibinfo {author} {\bibfnamefont
  {J.}~\bibnamefont {Zhou}}, \bibinfo {author} {\bibfnamefont {K.}~\bibnamefont
  {Han}}, \bibinfo {author} {\bibfnamefont {W.~B.}\ \bibnamefont {Wu}},
  \bibinfo {author} {\bibfnamefont {Y.~P.}\ \bibnamefont {Feng}}, \bibinfo
  {author} {\bibfnamefont {J.~M.~D.}\ \bibnamefont {Coey}}, \ and\ \bibinfo
  {author} {\bibfnamefont {T.}~\bibnamefont {Venkatesan}},\ }\href {\doibase
  10.1103/PhysRevB.90.125156} {\bibfield  {journal} {\bibinfo  {journal}
  {Physical Review B}\ }\textbf {\bibinfo {volume} {90}},\ \bibinfo {pages}
  {125156} (\bibinfo {year} {2014})}\BibitemShut {NoStop}%
\bibitem [{\citenamefont {Ohtomo}\ and\ \citenamefont
  {Hwang}(2004)}]{Ohtomo2004}%
  \BibitemOpen
  \bibfield  {author} {\bibinfo {author} {\bibfnamefont {A.}~\bibnamefont
  {Ohtomo}}\ and\ \bibinfo {author} {\bibfnamefont {H.~Y.}\ \bibnamefont
  {Hwang}},\ }\href {\doibase 10.1063/1.1668329} {\bibfield  {journal}
  {\bibinfo  {journal} {Applied Physics Letters}\ }\textbf {\bibinfo {volume}
  {84}},\ \bibinfo {pages} {1716} (\bibinfo {year} {2004})}\BibitemShut
  {NoStop}%
\bibitem [{\citenamefont {Yoon}\ \emph {et~al.}(shed)\citenamefont {Yoon},
  \citenamefont {Inoue}, \citenamefont {Swartz}, \citenamefont {Hikita},\ and\
  \citenamefont {Hwang}}]{Hyeokunpublished}%
  \BibitemOpen
  \bibfield  {author} {\bibinfo {author} {\bibfnamefont {H.}~\bibnamefont
  {Yoon}}, \bibinfo {author} {\bibfnamefont {H.}~\bibnamefont {Inoue}},
  \bibinfo {author} {\bibfnamefont {A.~G.}\ \bibnamefont {Swartz}}, \bibinfo
  {author} {\bibfnamefont {Y.}~\bibnamefont {Hikita}}, \ and\ \bibinfo {author}
  {\bibfnamefont {H.~Y.}\ \bibnamefont {Hwang}},\ }\href@noop {} {\  (\bibinfo
  {year} {unpublished})}\BibitemShut {NoStop}%
\bibitem [{\citenamefont {Chandra}\ \emph {et~al.}(1979)\citenamefont
  {Chandra}, \citenamefont {Wood}, \citenamefont {Woodard},\ and\ \citenamefont
  {Eastman}}]{Chandra1979}%
  \BibitemOpen
  \bibfield  {author} {\bibinfo {author} {\bibfnamefont {A.}~\bibnamefont
  {Chandra}}, \bibinfo {author} {\bibfnamefont {C.~E.~C.}\ \bibnamefont
  {Wood}}, \bibinfo {author} {\bibfnamefont {D.~W.}\ \bibnamefont {Woodard}}, \
  and\ \bibinfo {author} {\bibfnamefont {L.~F.}\ \bibnamefont {Eastman}},\
  }\href {\doibase 10.1016/0038-1101(79)90138-2} {\bibfield  {journal}
  {\bibinfo  {journal} {Solid State Electronics}\ }\textbf {\bibinfo {volume}
  {22}},\ \bibinfo {pages} {645} (\bibinfo {year} {1979})}\BibitemShut
  {NoStop}%
\bibitem [{\citenamefont {Neville}\ \emph {et~al.}(1972)\citenamefont
  {Neville}, \citenamefont {Hoeneisen},\ and\ \citenamefont
  {Mead}}]{Neville1972}%
  \BibitemOpen
  \bibfield  {author} {\bibinfo {author} {\bibfnamefont {R.~C.}\ \bibnamefont
  {Neville}}, \bibinfo {author} {\bibfnamefont {B.}~\bibnamefont {Hoeneisen}},
  \ and\ \bibinfo {author} {\bibfnamefont {C.~A.}\ \bibnamefont {Mead}},\
  }\href {\doibase 10.1063/1.1661463} {\bibfield  {journal} {\bibinfo
  {journal} {Journal of Applied Physics}\ }\textbf {\bibinfo {volume} {43}},\
  \bibinfo {pages} {2124} (\bibinfo {year} {1972})}\BibitemShut {NoStop}%
\end{thebibliography}%

\end{document}